\documentclass{article}

\usepackage[french,english]{babel} 
\def\english{\selectlanguage{english}}
\def\french{\selectlanguage{french}}
\usepackage{RR}
\usepackage{myNotation}
\usepackage{amssymb}
\usepackage{latexsym}

\RRNo{4082}
\RRdate{December 2000} 
\RRauthor{Pierre Deransart\thanks{\texttt{pierre.deransart@inria.fr}}
\and
Jan--Georg Smaus\thanks{ \texttt{jan.smaus@cwi.nl}, 
CWI, Kruislaan 413, 1098 SJ Amsterdam, The Netherlands.}}

\authorhead{Deransart \& Smaus} 

\RRetitle{Well-Typed Logic Programs Are not Wrong} 
\RRtitle{Les programmes logiques bien typ\'es ont tout bon}

\titlehead{Well-Typed Logic Programs Are not Wrong} 
\RRnote{This report is the complete version of a paper presented at
FLOPS 2001~\cite{DS01}. It contains all proofs omitted there for space 
reasons. Copies of this report obtained from INRIA contain a mistake
(the converse of Theorem~\ref{type-tree-theorem} is claimed to hold)
which is corrected in the present version.}

\RRresume{On \'etudie ici les syt\`emes de typage prescriptif (\`a la G\"odel ou
Mercury) pour la programmation en logique. Dans de tels syst\`emes, le
typage est {\em statique}, mais il garantit une propri\'et\'e
op\'erationnelle: si un programme est ``bien typ\'e'', alors tous les
buts obtenus par d\'erivation d'un but ``bien typ\'e'' sont eux-m\^eme
``bien typ\'es''. Le syst\`eme est alors dit {\em stable par
r\'eduction} (``subject reduction''). Nous montrons dans ce papier que
cette propri\'et\'e de stabilit\'e est en fait aussi {\em
d\'eclarative}.  Cette vue d\'eclarative nous conduit \`a
reconsid\'erer une condition habituellement admise et n\'ec\'essaire
pour la stabilit\'e par r\'eduction, dite {\em condition de t\^ete}
(``head condition''). Cette condition stipule que le type des t\^etes
des clauses d'un programme ``bien typ\'e'' doit \^etre une variante
(et non une instance quelconque) du type d\'eclar\'e de son
pr\'edicat. Il est alors possible de formuler des conditions plus
g\'en\'erales qui r\'etablissent une certaine sym\'etrie entre les
t\^etes et les atomes du corps, et qui garantissent en fait que dans
toutes d\'erivations les types des termes unifiables sont eux aussi
unifiables. On discute enfin les implications possibles d'un tel
r\'esultat, en particulier la relation entre la condition de t\^ete et 
la {\em recursion polymorphique}, un concept connu dans la
programmation fonctionnelle.
}

\RRabstract{We consider prescriptive type systems for logic programs (as in
G\"odel or Mercury). In such systems, the typing is {\em static}, but
it guarantees an operational property: if a program is ``well-typed'',
then all derivations starting in a ``well-typed'' query are again
``well-typed''. This property has been called {\em subject
reduction}. We show that this property can also be phrased as a
property of the {\em proof-theoretic} semantics of logic programs,
thus abstracting from the usual operational (top-down) semantics.
This proof-theoretic view leads us to questioning a condition
which is usually considered necessary for subject reduction, namely
the {\em head condition}. It states that the head of each clause must 
have a type which is a variant (and not a proper instance) of the 
declared type. We provide a more general condition, thus
reestablishing a certain symmetry between heads and body atoms. The
condition ensures that in a derivation, the types of two unified terms 
are themselves unifiable. We discuss possible implications of this result.
We also discuss the relationship between the head condition and 
{\em polymorphic recursion}, a concept known in functional programming.}
\RRmotcle{programmation en logique, syst\`eme de type (prescriptif), 
``subject reduction'', polymorphisme, condition de t\^ete, 
arbre de d\'erivation, recursion polymorphique}
\RRkeyword{Logic programming, (prescriptive) type system, subject
reduction, polymorphism, head condition, derivation tree,
polymorphic recursion}
\RRprojet{Contraintes} 
\RRtheme{2} 
\URRocq 

\newtheorem{theorem}{Theorem}{\bfseries}{\rm} 
\newtheorem{rmtheorem}[theorem]{Theorem}{\bfseries}{\rm} 
\newtheorem{rmlemma}[theorem]{Lemma}{\bfseries}{\rm} 
{\bfseries}{\rm} 
\newtheorem{propo}[theorem]{Proposition}{\bfseries}{\rm} 
\newtheorem{definition}{Definition}{\bfseries}{\rm} 
\newtheorem{defi}[definition]{Definition}{\bfseries}{\rm} 
\newtheorem{example}{Example}{\bfseries}{\rm} 
{\bfseries}{\rm} 

\newenvironment{proof}{\bfseries Proof:\rm} {\hfill$\square$\vspace{4mm}}

\renewcommand\ftau{f_{\tau_1 \dots \tau_m\rightarrow\tau}}
\renewcommand\ptau{p_{\tau_1 \dots \tau_m}}

\begin{document}

\makeRR   

\author{Pierre Deransart\inst{1} \and Jan-Georg Smaus\inst{2}}

\section{Introduction}
Prescriptive types are used in logic programming 
(and other paradigms)
to restrict the
underlying syntax so that only ``meaningful'' expressions are
allowed. This allows for many programming errors to be detected by the
compiler. Moreover, it ensures that once a program
has passed the compiler, the types of arguments of predicates can be
ignored at runtime, since it is guaranteed that they will be of
correct type.  This has been turned into the famous slogan~\cite{M78,MO84}
\begin{quote}
Well-typed programs cannot go wrong. 
\end{quote}
Adopting the terminology from the theory of the
$\lambda$-calculus~\cite{Tho91}, this property of a typed program is
called {\em subject reduction}.  For the simply typed
$\lambda$-calculus, subject reduction states that the type of a
$\lambda$-term is invariant under reduction. Translated to logic
programming, this means that resolving a ``well-typed'' query with a
``well-typed'' clause will always result in a ``well-typed'' query,
and so the successive queries obtained during a derivation are all
``well-typed''.

From this observation, it is clear that subject reduction is 
a property of the {\em operational} semantics of a
logic program, i.e., SLD resolution~\cite{L87}. 
In this paper, we show that it is also a property of
the proof-theoretic semantics based on {\em derivation
trees}. This is obtained by showing that using ``well-typed'' clauses,
only ``well-typed'' derivation trees can be constructed, giving rise
to the new slogan: 
\begin{quote}
Well-typed programs {\em are} not wrong. 
\end{quote}
The {\em head condition}, which is a condition on the program
(clauses)~\cite{HT92-new}, is usually considered to be crucial for
subject reduction. The second objective of this paper is to analyse
the head condition in this new light and open the field for
generalisations, of which we introduce one.

The head condition, also called 
{\em definitional genericity}~\cite{LR91},  states that the types of the arguments of a clause
head must be a variant\footnote{A variant is obtained by renaming the
type parameters in a type.}  (and not a proper instance) of the
declared type of the head predicate. This
condition 
imposes a distinction between ``definitional'' occurrences (clause
heads) and ``applied'' occurrences (body atoms) of a predicate. In
contrast, the proof-theoretic view of subject reduction we propose
reestablishes a certain symmetry between the different occurrences. By
this generalisation, the class of programs for which subject reduction
is guaranteed is enlarged.

This paper is organised as follows. Section \ref{prelim-sec} contains some
preliminaries. Section~\ref{trees-sec} introduces our proof-theoretic
notion of subject reduction. 
Section~\ref{conditions-sec} gives conditions for subject reduction,
and in particular, a generalisation of the head condition.
In Section~\ref{is-restrictive-sec}, we discuss, in the light of these
results, the usefulness of the head condition and its generalisation.
We also exhibit an interesting relationship between
the head condition and {\em polymorphic recursion}~\cite{KFU93-short}.
Section~\ref{conclusion-sec} concludes by mentioning possible
applications of these results. 

\section{Preliminaries}\label{prelim-sec}

We assume familiarity with the standard concepts of logic
programming~\cite{L87}. 
To simplify the notation, a vector such as $o_1,\dots,o_m$ 
is often denoted by $\bar{o}$. The restriction of a 
substitution $\theta$ to the variables in a syntactic object $o$ is
denoted as $\theta \restr{o}$, and analogously for type 
substitutions (see Subsec.~\ref{typed-prog-subsec}).
The relation symbol of an atom $a$ is denoted by $Rel(a)$.

When we refer to a {\em clause in a program}, we usually mean a copy
of this clause whose variables are renamed apart from variables
occurring in other objects in the context. A query is a sequence of
atoms. A 
query $Q'$ is {\bf derived from} a query $Q$, denoted $Q \leadsto Q'$,
if
$Q= a_1,\dots,a_m$, 
$Q'=(a_1,\dots,a_{k-1},B,a_{k+1},\dots,a_m)\theta$, 
and $h\leftarrow B$ is a clause (in a program usually clear from  
the context) such that $h$ and $a_k$ are unifiable with MGU $\theta$.
A {\bf derivation} $Q \leadsto^* Q'$ is defined in the usual way.
Given a program $P$, the {\bf immediate consequence operator} 
$T_P$ is defined by
$
T_P(M) =
\{h\theta \mid 
h\leftarrow a_1,\dots,a_m \in P,\;
a_1\theta,\dots,a_m\theta \in M\}$.

\subsection{Derivation Trees}
A key element of this work is the proof-theoretic semantics of
logic programs based on derivation trees~\cite{DM93}.
We recall some important notions and basic 
results.

\begin{defi}
\label{instance-name-def}
An {\bf instance name} of a clause $C$ is a pair of the form 
$\langle C, \theta \rangle$, where $\theta$ is a substitution.
\end{defi}

\begin{defi}
\label{der-tree-def}
Let $P$ be a program. A {\bf derivation tree} for $P$ is a labelled
ordered tree~\cite{DM93} such that:
\begin{enumerate}
\item
Each leaf node is labelled by $\bot$ or an instance name 
$\langle C,\theta \rangle$ of a clause\footnote{
Recall that $C$ is renamed apart from any other clause in the
same tree.}  in $P$; 
each non-leaf node is labelled by an instance name 
$\langle C,\theta \rangle$ of a clause in $P$.
\item 
If a node is labelled by 
$\langle h\leftarrow a_1,\dots,a_m,\theta \rangle$, where $m \geq 0$,
then this node has $m$ children, and for $i \in \onetom$, the $i$th
child is labelled either $\bot$, or 
$\langle h'\leftarrow B,\theta' \rangle$ where $h'\theta'=a_i\theta$.
\end{enumerate}
Nodes labelled $\bot$ are {\bf incomplete}, all other nodes are 
{\bf complete}.
A derivation tree containing only complete nodes is a 
{\bf proof tree}. 
\end{defi}

To define the semantics of logic programs, it is useful to associate
an atom with each node in a derivation tree in the following way.

\begin{defi}
\label{node-atom-def}
Let $T$ be a derivation tree. For each node $n$ in $T$, the 
{\bf node atom} of $n$, denoted $atom(n)$, is defined as follows: 
If $n$ is labelled
$\langle h\leftarrow B,\theta \rangle$, then
$h \theta$ is the node atom of $n$; if $n$ is labelled
$\bot$, and $n$ is the 
$i$th child of its parent labelled 
$\langle h\leftarrow a_1,\dots,a_m,\theta \rangle$, then
$a_i\theta$ is the node atom of $n$. 
If $n$ is the root of $T$ then $atom(n)$ is the {\bf head of $T$},
denoted $head(T)$.
\end{defi}

Derivation trees are obtained by grafting instances of clauses of a
program. To describe this construction in a general way, we
define the following concept.

\begin{defi}
\label{skel-def}
Let $P$ be a program. A {\bf skeleton (tree)} for $P$ is a labelled
ordered tree such that:
\begin{enumerate}
\item
Each leaf node is labelled by $\bot$ or a clause
in $P$, and
each non-leaf node is labelled by a clause in $P$.
\item 
If a node is labelled by 
$h\leftarrow a_1,\dots,a_m$, where $m \geq 0$, then
this node has $m$ children, and for $i \in \onetom$, the $i$th child is
labelled either $\bot$, or 
$h'\leftarrow B$ where $Rel(h') = Rel(a_i)$. 
\end{enumerate}
\end{defi}

The {\bf skeleton of a tree} $T$, denoted $Sk(T)$, is the skeleton
obtained from $T$ by replacing each label 
$\langle C, \theta \rangle$ with $C$. Conversely, we
say that $T$ is a {\bf derivation tree based on} $Sk(T)$.

\begin{defi}
\label{eqs-def}
Let $S$ be a skeleton. We define
\[
\begin{array}{ll}
\mathit{Eq}(S) =
\{
a_i=h'
\mid &
\mbox{there exist complete nodes $n$, $n'$ in $S$ 
such that}\\ 
&
\mbox{$\bullet$ $n'$ is the $i$th child of $n$,}\\
&
\mbox{$\bullet$ $n$ is labelled $h\leftarrow a_1,\dots,a_m$,}\\
&
\mbox{$\bullet$ $n'$ is labelled $h'\leftarrow B$}
\}
\end{array}
\]
Abusing notation, we frequently identify the set of equations with the
conjunction or sequence of all equations contained in it.
If $\mathit{Eq}(S)$ has a unifier then we call $S$ a {\bf proper} skeleton.
\end{defi}

\begin{propo}\cite[Prop.~2.1]{DM93}\label{der-tree-exists-prop}
Let $S$ be a skeleton. A derivation tree based on $S$ exists if and
only if $S$ is proper.
\end{propo}

\begin{rmtheorem}\cite[Thm.~2.1]{DM93}
Let $S$ be a skeleton and $\theta$ an MGU of $\mathit{Eq}(S)$. Let $D(S)$ be
the tree obtained from $S$ by replacing each node label $C$ with the
pair $\langle C, \theta\restr{C} \rangle$. Then $D(S)$ is a most 
general derivation tree based on $S$ (i.e., any other derivation
tree based on $S$ is an instance of $D(S)$).
\end{rmtheorem}

\begin{example}\label{proof-tree-ex}
Figure~\ref{no-pure-BC-fig} shows a program, one of 
its derivation trees,  and the skeleton of the derivation tree.
\end{example}

\setlength{\unitlength}{0.46cm}
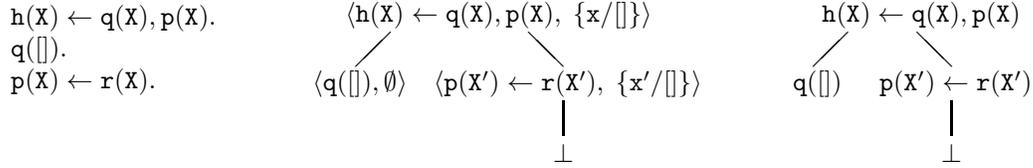
\begin{figure}[t]
\begin{picture}(6,5)
\put(0,3.3){
\begin{minipage}{6cm}
{
$\tt h(X) \leftarrow q(X), p(X).$\\
$\tt q([]).$\\
$\tt p(X) \leftarrow r(X).$
}
\end{minipage}}
\end{picture}
\hfill
\begin{picture}(12,5)
\put(4,0){\makebox(8,1){$\bot$}}
\put(8,1){\line(0,1){1}}
\put(0,2){\makebox(4,1){
  $\langle \tt q([]),\emptyset \rangle$}}
\put(4,2){\makebox(8,1){
  $\langle \tt p(X') \leftarrow r(X'),\; 
  \{x'/[]\}\rangle$}}
\put(2,3){\line(1,1){1}}
\put(8,3){\line(-1,1){1}}
\put(1,4){\makebox(10,1){
  $\langle \tt h(X) \leftarrow q(X) \und p(X),\;
  \{x/[]\}\rangle$}}
\end{picture}
\hfill
\begin{picture}(8,5)
\put(2,0){\makebox(6,1){$\bot$}}
\put(5,1){\line(0,1){1}}
\put(0,2){\makebox(2,1){
  $\tt q([])$}}
\put(2,2){\makebox(6,1){
  $\tt p(X') \leftarrow r(X')$}}
\put(1,3){\line(1,1){1}}
\put(5,3){\line(-1,1){1}}
\put(0,4){\makebox(8,1){
  $\tt h(X) \leftarrow q(X), p(X)$}}
\end{picture}
\caption{A program, a derivation tree and its skeleton
\label{no-pure-BC-fig}\label{trees-fig}}
\end{figure}

To model derivations for a program $P$ and a query
$Q$, we assume that $P$ contains an
additional clause ${\tt go} \leftarrow Q$,
 where $\tt go$ is a new predicate symbol. 

We recall the following straightforward
correspondences between derivations, the $T_P$-semantics and
derivation trees. 

\begin{propo}\label{semant-equiv-propo}
Let $P$ be a program. Then 
\begin{enumerate}
\item \label{tree=TP}
$a \in \mathit{lfp}(T_P)$ if and only if $a=head(T)$ for some proof tree
$T$ for $P$,
\item \label{tree=operational}
$Q \leadsto^* Q'$ if and only if $Q'$ is the 
sequence of node atoms of incomplete nodes 
of a most general derivation tree for 
$P \cup \{{\tt go} \leftarrow Q\}$ with head $\tt go$, visited
left to right.
\end{enumerate}
\end{propo}

\subsection{Typed Logic Programming}\label{typed-prog-subsec}
We assume a type system for logic programs with parametric
polymorphism but without subtyping, 
as realised in the languages G\"odel~\cite{goedel} or
Mercury~\cite{mercury}. 

The set of types $\mathcal T$ is given by the term structure based on 
a finite set of {\bf constructors} $\mathcal K$, where with each 
$K\in\mathcal{K}$ an arity 
$m\geq 0$ is associated (by writing $K/m$), 
and a denumerable set $\mathcal U$ of 
{\bf parameters}. 
A {\bf type substitution} is an idempotent mapping
from parameters to types which is the identity almost everywhere.
The set of parameters in a syntactic object $o$ is denoted by $\pars(o)$.

We assume a denumerable set $\mathcal V$ of {\bf variables}.
The set of variables in a syntactic object $o$ is
denoted by $\vars(o)$.
A {\bf variable typing} is a mapping from a finite subset of 
$\mathcal V$ to $\mathcal T$, written as
$\{x_1:\tau_1,\dots,x_m:\tau_m\}$. 

We assume a finite set $\mathcal F$ (resp.~$\mathcal P$) 
of {\bf function} (resp.~{\bf predicate}) symbols, each
with an arity and a {\bf declared type} associated with it, 
such that:
for each $f \in \mathcal F$, the declared type has the form
$(\tau_1,\dots,\tau_m,\tau)$, 
where $m$ is the arity of $f$,
$(\tau_1,\dots,\tau_m)\in {\mathcal T}^m$,
and $\tau$ 
satisfies the {\em transparency condition} \cite{HT92-new}:\label{transparency}
$\pars(\tau_1,\dots,\tau_m) \subseteq \pars(\tau)$;
for each $p \in \mathcal P$, the declared type has the form
$(\tau_1,\dots,\tau_m)$, 
where $m$ is the arity of $p$ and
$(\tau_1,\dots,\tau_m)\in {\mathcal T}^m$. 
We often indicate the declared types by writing 
$\ftau$ and $\ptau$, however we assume that the parameters in
$\tau_1,\dots,\tau_m,\tau$ are fresh for each occurrence of $f$ or $p$.
We assume that there is a special predicate symbol 
$=_{u,u}$
where $u\in \mathcal U$.

Throughout this paper, we assume $\mathcal K$, $\mathcal F$, and 
$\mathcal P$ arbitrary but fixed.
The {\bf typed language}, i.e.\ a language of terms, atoms etc.\ based
on $\mathcal K$, $\mathcal F$, and $\mathcal P$, is defined by the
rules
in Table \ref{rules-tab}. All objects are defined relative to 
a variable typing $U$, and $\_\vdash\dots$ stands for ``there exists
$U$ such that $U\vdash\dots$''. The expressions below the line are called 
{\bf type judgements}.

\begin{table}[t]
\caption{Rules defining a typed language\label{rules-tab}}
\begin{center}
\begin{tabular}{lll}
{\em (Var)} &
$\{x:\tau,\dots\}\vdash x:\tau$\\[2ex]
{\em (Func)} &
\Large
$\frac%
  {U\vdash t_1:\tau_1\Theta\ \cdots \ U\vdash t_m:\tau_m\Theta}%
  {U\vdash\ftau(t_1,\dots,t_m):\tau\Theta}$ &
$\Theta$ is a type substitution\\[2ex]
{\em (Atom)} &
\Large
$\frac%
  {U\vdash t_1:\tau_1\Theta\ \cdots \ U\vdash t_m:\tau_m\Theta}%
  {U\vdash\ptau(t_1,\dots,t_m)\; \mathit{Atom}}$ &
$\Theta$ is a type substitution\\[2ex]
{\em (Query)} &
\Large
$\frac%
  {U\vdash A_1\; \mathit{Atom}\ \cdots \  U\vdash A_m\; \mathit{Atom}}%
  {U\vdash A_1,\dots,A_m\; \mathit{Query}}$ \\[2ex]
{\em (Clause)} & 
\Large
$\frac%
  {U\vdash A\; \mathit{Atom} \quad U\vdash Q\; Query}%
  {U \vdash A \leftarrow Q\; \mathit{Clause}}$ \\[2ex]
{\em (Program)} &
\Large
$\frac%
  {\_\vdash C_1\; \mathit{Clause} \ \cdots \ \_\vdash C_m\; \mathit{Clause}}%
  {\_\vdash \{C_1,\dots,C_m\}\; \mathit{Program}}$ \\[2ex]
{\em (Queryset)} &
\Large
$\frac%
  {\_\vdash Q_1\; \mathit{Query}\ \cdots \  \_\vdash Q_m\; \mathit{Query}}%
  {\_\vdash \{Q_1,\dots, Q_m\} \; \mathit{Queryset}}$ 
\end{tabular}
\end{center}
\end{table}

Formally, a proof of a type judgement is a tree where the nodes are 
labelled with judgements and the edges are labelled with rules (e.g.\
see Fig.\ \ref{judgment-fig})~\cite{Tho91}. 
From the form of the rules, it is clear that in order to prove
any type judgement, we must, for each occurrence of a term $t$ in the
judgement, prove a judgement $\dots \vdash t:\tau$ for some $\tau$.
We now define the most general such $\tau$. It exists and can be
computed by {\em type inferencing algorithms}~\cite{B95}. 

\begin{figure}[t]
\begin{center}
$
\begin{array}{ccccc}
  & \hspace{0.5em} & \vdots & & \vdots 
\\ 
  \raisebox{0pt}[0pt]{\vdots} & & 
  U\vdash \bar{t}_1:\bar{\tau}_1  & 
  \raisebox{-1.3ex}{\dots} &
  U\vdash \bar{t}_m:\bar{\tau}_m 
\\
  \cline{3-3}\cline{5-5}
  U\vdash \bar{t}:\bar{\tau} & & 
  U\vdash p_1(\bar{t}_1)\ \mathit{Atom} & &
  U\vdash p_m(\bar{t}_m)\ \mathit{Atom}
\\
  \cline{1-1}\cline{3-5}
  U\vdash p(\bar{t})\ \mathit{Atom} & &  
  \multicolumn{3}{c}{U\vdash p_1(\bar{t}_1),\dots,p_m(\bar{t}_m)\ \mathit{Query}}
\\\hline
  \multicolumn{5}{c}{U\vdash p(\bar{t})\leftarrow p_1(\bar{t}_1),\dots,p_m(\bar{t}_m)\ \mathit{Clause}}
\end{array}
$
\end{center}
\caption{Proving a type judgement\label{judgment-fig}}
\end{figure}

\begin{defi}\label{most-general-def}
Consider a judgement
$U\vdash p(\bar{t})\leftarrow p_1(\bar{t}_1),\dots,p_m(\bar{t}_m)\ 
\mathit{Clause}$, 
and a proof of this judgement containing judgements
$U\vdash \bar{t}:\bar{\tau}$, 
$U\vdash \bar{t}_1:\bar{\tau}_1$, \dots,
$U\vdash \bar{t}_m:\bar{\tau}_m$ (see Fig.\ \ref{judgment-fig})
such that 
$(\bar{\tau},\bar{\tau}_1,\dots,\bar{\tau}_m)$ is 
most general (wrt.\ all such proofs). 
We call $(\bar{\tau},\bar{\tau}_1,\dots,\bar{\tau}_m)$ the 
{\bf most general type}  
of $p(\bar{t})\leftarrow p_1(\bar{t}_1),\dots,p_m(\bar{t}_m)$ 
{\bf wrt.\ $U$}.

Moreover, consider the variable typing $U'$ and the proof of the judgement
$U'\vdash p(\bar{t})\leftarrow p_1(\bar{t}_1),\dots,p_m(\bar{t}_m)\ 
\mathit{Clause}$\hspace{0.5em} containing judgments
$U'\vdash \bar{t}:\bar{\tau}$, 
$U'\vdash \bar{t}_1:\bar{\tau}_1$, \dots,
$U'\vdash \bar{t}_m:\bar{\tau}_m$ such that 
$(\bar{\tau},\bar{\tau}_1,\dots,\bar{\tau}_m)$ is 
most general (wrt.\ all such proofs and all possible $U'$).
We call $(\bar{\tau},\bar{\tau}_1,\dots,\bar{\tau}_m)$ the 
{\bf most general type} 
of $p(\bar{t})\leftarrow p_1(\bar{t}_1),\dots,p_m(\bar{t}_m)$.
\end{defi}

The following example explains the difference between the most general 
type wrt.\ a fixed variable typing, and the most general type as such.

\begin{example}\label{most-general-example}
Consider function 
$\tt nil_{\rightarrow list(U)}$ and clause 
$C = \tt p \leftarrow X\! =\! nil,$ 
$\tt nil\! =\! nil$.  
Fixing $U= \{\tt X:list(int)\}$, the judgement
$U\vdash C\ \mathit{Clause}$ can be proven using 
the judgements
$U\vdash \tt X: list(int)$ and then 
$U\vdash \tt nil: list(int)$ for {\em each} occurrence 
of $\tt nil$. It can also be proven 
using the judgements
$U\vdash \tt X: list(int)$ and then  
$U\vdash \tt nil: list(int)$ (for the first occurrence of 
$\tt nil$) and then
$U\vdash \tt nil: list(V)$ (for the other two occurrences of 
$\tt nil$). In the latter case, we obtain 
$(\tt list(int), list(int), list(V), list(V))$, the most general
type of $C$ wrt.\ $U$. 
Moreover, 
$(\tt list(V'), list(V'), list(V), list(V))$ is the most general type
of $C$ (choose $U' = \{\tt X:list(V')\}$).
\end{example}

\begin{defi}\label{substitution-def}
If 
$U\vdash x_1\!=\!t_1, \dots, x_m\! =\! t_m\ Query$
where $x_1,\dots,x_m$ are distinct variables and for
each $i\in \onetom$, $t_i$ is a term distinct from $x_i$, then 
$(\{ x_1/t_1,\dots,x_m/t_m\}, U)$ is a 
{\bf typed (term) substitution}. 
\end{defi}

We shall need three fundamental lemmas introduced
in~\cite{HT92-new}.\footnote{
Note that some results in~\cite{HT92-new}
have been shown to be faulty (Lemmas 1.1.7, 1.1.10 and 1.2.7), 
although we believe that these mistakes only affect type systems 
which include subtyping.}

\begin{rmlemma}~\cite[Lemma~1.2.8]{HT92-new}\label{variable-typing-lemma}
Let $U$ be a variable typing and $\Theta$ a type
substitution.
If $U\vdash t:\sigma$, then 
$U\Theta\vdash t:\sigma\Theta$.
Moreover, if 
$U\vdash A\ \mathit{Atom}$ then $U\Theta\vdash A\ \mathit{Atom}$, and 
likewise for queries and clauses.
\end{rmlemma}
\begin{proof}
The proof is by structural induction.  For the base case, suppose
$U\vdash x:\sigma$ where $x\in \mathcal V$. Then 
$x: \sigma \in U$ and hence $x:\sigma\Theta \in U\Theta$. Thus
$U\Theta\vdash x:\sigma\Theta$.

Now consider 
$U\vdash \ftau(t_1,\dots,t_m):\sigma$ 
where the inductive hypothesis holds for $t_1,\dots,t_m$. 
By Rule {\em (Func)}, there exists a type 
substitution $\Theta'$ such that $\sigma=\tau\Theta'$ 
and $U\vdash t_i:\tau_i\Theta'$ 
for each $i \in \onetom$. 
By the inductive hypothesis, 
$U\Theta \vdash t_i:\tau_i\Theta'\Theta$ for each
$i\in\onetom$, and hence by 
Rule {\em (Func)},
$U\Theta\vdash \ftau(t_1,\dots,t_m): \tau\Theta'\Theta$.

The rest of the proof is now trivial.
\end{proof}

\begin{rmlemma}~\cite[Lemma~1.4.2]{HT92-new}\label{apply-substitution-lemma}
Let $(\theta,U)$ be a typed substitution.
If $U\vdash t:\sigma$ then $U\vdash t\theta :\sigma$. Moreover, if 
$U\vdash A\ \mathit{Atom}$ then $U\vdash A\theta\ \mathit{Atom}$,
and likewise for queries and clauses.
\end{rmlemma}
\begin{proof}
The proof is by structural induction. For the base case, suppose
$U\vdash x:\sigma$ where $x\in \mathcal V$. If 
$x\theta=x$, there is nothing to show. If $x/t \in \theta$, then by
definition of a typed substitution, 
$U\vdash t: \sigma$.

Now consider 
$U\vdash \ftau(t_1,\dots,t_m) :\sigma$ 
where the inductive hypothesis holds for $t_1,\dots,t_m$.
By Rule {\em (Func)}, there exists a type
substitution $\Theta'$ such that $\sigma=\tau\Theta'$, 
and $U\vdash t_i:\tau_i\Theta'$ for each $i \in \onetom$. 
By the inductive hypothesis, 
$U\vdash t_i\theta:\tau_i\Theta'$ for each $i\in\onetom$, and hence by 
Rule {\em (Func)},
$U\vdash \ftau(t_1,\dots,t_m)\theta : \tau\Theta'$.

The rest of the proof is now trivial.
\end{proof}

\begin{rmlemma}~\cite[Thm.\ 1.4.1]{HT92-new}\label{is-substitution-lemma}
Let $E$ be a set (conjunction) of equations such that for some
variable typing $U$, we have $U\vdash E\ \mathit{Query}$. Suppose
$\theta$ is an MGU of $E$. Then $(\theta,U)$ is a typed substitution.
\end{rmlemma}
\begin{proof}
We show that the result is true when $\theta$ is computed using the
well-known Martelli-Montanari algorithm~\cite{MM82} which works by
transforming a set of equations $E=E_0$ into a set of the form required
in the definition of a typed substitution. Only the following two
transformations are considered here. The others are trivial.
\begin{enumerate}
\item\label{replacement}
If $x=t\in E_k$ and $x$ does not occur in $t$, 
then replace all occurrences of $x$ in all other equations in $E$ with 
$t$, to obtain $E_{k+1}$.
\item\label{decomposition}
If $f(t_1,\dots,t_m)=f(s_1,\dots,s_m)\in E_k$, then replace this
equation with $t_1=s_1,\dots,t_m=s_m$, to obtain $E_{k+1}$.
\end{enumerate}
We show that if $U\vdash E_k\ \mathit{Query}$ and
 $E_{k+1}$ is obtained by either of the above transformations,
 then $U\vdash E_{k+1}\ \mathit{Query}$.
For (\ref{replacement}), this follows from
Lemma~\ref{apply-substitution-lemma}. 

For (\ref{decomposition}), suppose $U\vdash E_k\ \mathit{Query}$ and
$f(t_1,\dots,t_m)=f(s_1,\dots,s_m)\in E_k$ where $f=\ftau$. 
By Rule {\em (Query)}, we must have
$U\vdash f(t_1,\dots,t_m)=_{u, u}f(s_1,\dots,s_m)\ \mathit{Atom}$,
and hence by Rule {\em (Atom)},
$U\vdash f(t_1,\dots,t_m): u\Theta$ and
$U\vdash f(s_1,\dots,s_m): u\Theta$ for some type substitution
$\Theta$. On the other hand, by Rule {\em (Func)}, 
$u\Theta= \tau\Theta_t$ and 
$u\Theta= \tau\Theta_s$ for some type substitutions $\Theta_s$ and
$\Theta_t$, and moreover for each $i\in\onetom$, we have
$U\vdash t_i: \tau_i\Theta_t$ and 
$U\vdash s_i: \tau_i\Theta_s$.
Since
$\pars(\tau_i) \subseteq \pars(\tau)$, it follows that 
$\tau_i\Theta_t = \tau_i\Theta_s$.\footnote{
Note how the transparency condition is essential to ensure that
subarguments in corresponding positions have identical types. This
condition was ignored in~\cite{MO84}.}
Therefore $U\vdash t_i=s_i\ \mathit{Atom}$, and so
$U\vdash E_{k+1}\ \mathit{Query}$.
\end{proof}

\section{Subject Reduction for Derivation Trees}
\label{trees-sec}

We first define subject reduction as 
a property of derivation trees and show that
it is equivalent to the usual operational notion.
We then show that a sufficient condition for 
subject reduction is that the types of all
unified terms are themselves unifiable.

\subsection{Proof-Theoretic and Operational Subject Reduction}
Subject reduction is a well-understood concept, yet it has to be
defined formally for each system.  We now provide two fundamental
definitions.

\begin{defi}
\label{subj-red-def}\label{subj-red-oper-def}
Let $\_\vdash P \ \mathit{Program}$ and 
$\_\vdash\mathcal Q\ \mathit{Queryset}$.
We say $P$ has 
{\bf (proof-theoretic) subject reduction wrt.~$\mathcal Q$} if
for every $ Q \in \mathcal Q$, for every
most general derivation tree $T$ for 
$P\cup \{{\tt go} \leftarrow Q\}$ with head $\tt go$, 
there exists a variable typing $U'$ such that for each 
node atom $a$ of $T$, 
$U'\vdash a\ \mathit{Atom}$.

$P$ has {\bf operational subject reduction wrt.~$\mathcal Q$} 
if for every $ Q \in \mathcal Q$, for every
 derivation $Q \leadsto^* Q'$ of $P$, we have
$\_\vdash Q'\ \mathit{Query}$.
\end{defi}

The reference to $\mathcal Q$ is omitted if 
$\mathcal Q=\{ Q\mid \_\vdash Q\ \mathit{Query}\}$.
The following theorem states a certain equivalence between the two
notions.

\begin{rmtheorem}\label{decl-oper-theo}
Let $\_\vdash P \ \mathit{Program}$ and $\_\vdash\mathcal Q\ \mathit{Queryset}$.
If $P$ has subject reduction wrt.~$\mathcal Q$, then 
$P$ has operational subject reduction wrt.~$\mathcal Q$.
If $P$ has operational subject reduction, then 
$P$ has subject reduction.
\end{rmtheorem}
\begin{proof}
The first statement is a straightforward consequence of
Prop.~\ref{semant-equiv-propo} (\ref{tree=operational}).

For the second statement, assume $U\vdash Q\ \mathit{Query}$, let
$\xi = Q \leadsto^* Q'$, and $T$ be the derivation tree
for $P \cup \{{\tt go} \leftarrow Q\}$ corresponding to $\xi$ (by
Prop.~\ref{semant-equiv-propo} (\ref{tree=operational})).

By hypothesis, there exists a variable typing $U'$ such that for each 
{\em incomplete} node $n$ of $T$, we have 
$U'\vdash atom(n)\ \mathit{Atom}$. To show that this also holds for {\em complete}
nodes, we transform $\xi$ into a derivation which ``records the
entire tree $T$''. This is done as follows: Let $\tilde{P}$ be the program
obtained from $P$ by replacing each clause $h \leftarrow B$ with
$h \leftarrow B, B$. Let us call the atoms in the second occurrence of 
$B$ {\em unresolvable}. Clearly 
$\_\vdash h \leftarrow B, B\ \mathit{Clause}$ for each such clause.

By induction on the length of derivations, one can show that $\tilde{P}$
has operational subject reduction. For a single derivation step, this
follows from the operational subject reduction of $P$.

Now let $\tilde{\xi}= {\tt go} \leadsto \tilde{Q}'$ be the
derivation for 
$\tilde{P} \cup \{{\tt go} \leftarrow Q, Q\}$ using in each
step the clause corresponding to the clause used in $\xi$ for that
step, and resolving only the resolvable atoms. First note that
since $\tilde{P}$ has operational subject reduction, there exists a
variable typing $U'$ such that $U'\vdash \tilde{Q}' \ \mathit{Query}$.
Moreover, since the unresolvable atoms are not
resolved in $\tilde{\xi}$, it follows that $\tilde{Q}'$ contains 
exactly the non-root node atoms of $T$. This however shows
that for each node atom $a$ of $T$, we have 
$U'\vdash a\ \mathit{Atom}$. Since the choice of $Q$ was arbitrary, 
$P$ has subject reduction.  
\end{proof}

The following example shows that in the second statement of the above
theorem, it is crucial that $P$ has operational subject reduction 
wrt.~{\em all} queries.

\begin{example}\label{oper-SR-only-ex}
Let 
$\mathcal{K} = \{\mathtt{list}/1, \mathtt{int}/0\}$,
$\mathcal{F} = \{
\mathtt{nil_{\rightarrow list(U)}},$
$\mathtt{cons_{U,list(U)\rightarrow list(U)}},$ 
$\mathtt{-1_{\rightarrow int}}, $
$\mathtt{0_{\rightarrow int}},\dots\}$, 
$\mathcal{P} = \{
\mathtt{p_{list(int)}},$ 
$\mathtt{r_{list(U)}}\}$, and $P$ be 
 
{
\begin{verbatim}
   p(X) <- r(X).                     r([X]) <- r(X).
\end{verbatim}
}

\noindent
For each derivation 
${\tt p(X)} \leadsto^* Q'_0$, we have
$Q'_0= \tt p(Y)$ or $Q'_0= \tt r(Y)$ for some 
${\tt Y}\in\mathcal{V}$, and so 
$\{\tt Y : list(int)\}\vdash {\tt p(Y)}\ \mathit{Query}$ or
$\{\tt Y : list(U)\}\vdash {\tt r(Y)}\ \mathit{Query}$.
Therefore $P$ has operational subject reduction 
wrt.~$\{\tt p(X)\}$. Yet the derivation trees for $P$ have heads
$\tt p(Y)$, $\tt p([Y])$, $\tt p([[Y]])$ etc., and
$\_\not\vdash {\tt p([[Y]])}\ \mathit{Query}$.
\end{example}

\subsection{Unifiability of Types and Subject Reduction}
We now lift the notion of skeleton to the type level.

\begin{defi}
\label{type-skel-def}
Let $\_\vdash P \ \mathit{Program}$ and 
$S$ be a skeleton for $P$. 
The {\bf type skeleton corresponding to} $S$
is a tree obtained from $S$ by replacing each 
node label 
$C_n = p(\bar{t})\leftarrow p_1(\bar{t}_1),\dots,p_m(\bar{t}_m)$
with 
$p(\bar{\tau})\leftarrow p_1(\bar{\tau}_1),\dots,p_m(\bar{\tau}_m)$,
where $(\bar{\tau},\bar{\tau}_1,\dots,\bar{\tau}_m)$ is the most
general type of $C_n$.\footnote{
Recall that the variables in $C_n$ and the parameters in 
$\bar{\tau},\bar{\tau}_1,\dots,\bar{\tau}_m$
are renamed apart from other node labels in the same (type) skeleton.} 
For a type skeleton $\mathit{TS}$, the {\bf type equation set}
$\mathit{Eq}(\mathit{TS})$ and a {\bf proper} type skeleton are defined as 
in Def.~\ref{eqs-def}.
\end{defi}

The following theorem states that subject reduction is ensured if
terms are unified only if their types are also unifiable. 

\begin{rmtheorem}\label{type-tree-theorem}
Let $\_\vdash P \ \mathit{Program}$ and $\_\vdash \mathcal{Q} \ \mathit{Queryset}$.
$P$ has subject reduction wrt.~$\mathcal Q$  if for each proper
skeleton $S$ of $P\cup\{{\tt go} \leftarrow Q\}$ with head $\tt go$, 
where $Q\in\mathcal{Q}$, 
the {\em type} skeleton corresponding to $S$ is proper.
\end{rmtheorem}
\begin{proof}
Let $S$ be an arbitrary proper
skeleton for 
$P\cup\{{\tt go}\leftarrow Q\}$ with head $\tt go$, where 
$Q\in \mathcal Q$. Let $\theta=MGU(Eq(S))$ and
$\Theta=MGU(Eq(TS))$. 
For each node $n$ in $S$, labelled 
$p(\bar{t})\leftarrow p_1(\bar{t}_1),\dots,p_m(\bar{t}_m)$ in
$S$ and
$p(\bar{\tau})\leftarrow p_1(\bar{\tau}_1),\dots,p_m(\bar{\tau}_m)$ in 
$TS$, let $U_n$ be the variable typing such that
$U_n\vdash 
(\bar{t},\bar{t}_1,\dots,\bar{t}_m):
(\bar{\tau},\bar{\tau}_1,\dots,\bar{\tau}_m)$.
Let
\[
U = 
\bigcup_{n \in S}
U_n\Theta.
\]
Consider a pair of nodes $n$, $n'$ in $S$ such that 
$n'$ is a child of $n$, and the equation
$p(\bar{s})=p(\bar{s}') \in Eq(S)$ corresponding to this pair (see
Def.~\ref{eqs-def}). Consider also the equation
$p(\bar{\sigma}) = p(\bar{\sigma}') \in Eq(TS)$ corresponding to the pair 
$n$, $n'$ in $TS$. Note that 
$U_n   \vdash \bar{s}: \bar{\sigma}$ and 
$U_{n'}\vdash \bar{s}':\bar{\sigma}'$.
By Lemma~\ref{variable-typing-lemma}, 
$U\vdash\bar{s} :\bar{\sigma} \Theta$ and 
$U\vdash\bar{s}':\bar{\sigma}'\Theta$.
Moreover, since $\Theta=MGU(Eq(TS))$, we have 
$\bar{\sigma}\Theta=\bar{\sigma}'\Theta$.
 Therefore 
$U\vdash p(\bar{s}) = p(\bar{s}')\ \mathit{Atom}$. 
Since the same reasoning applies for any equation in
$Eq(S)$, by Lemma~\ref{is-substitution-lemma}, $(\theta,U)$ is a typed 
substitution.

Consider a node $n''$ in $S$ with node atom $a$.
 Since $U_{n''}\vdash  a\ \mathit{Atom}$, 
by Lemma~\ref{variable-typing-lemma}, 
$U\vdash a\ \mathit{Atom}$. and
by Lemma~\ref{apply-substitution-lemma},
$U\vdash a\theta\ \mathit{Atom}$.
Therefore $P$ has subject
reduction wrt.~$\mathcal Q$. 
\end{proof}

\setlength{\unitlength}{0.42cm}
\begin{figure}[t]
\begin{picture}(12,7)
\put(0,0){\makebox(12,1){$\tt r([X''']) \leftarrow r(X''')$}}
\put(6,1){\line(0,1){1}}
\put(0,2){\makebox(12,1){$\tt r([X'']) \leftarrow r(X'')$}}
\put(6,3){\line(0,1){1}}
\put(0,4){\makebox(12,1){$\tt p(X') \leftarrow r(X')$}}
\put(6,5){\line(0,1){1}}
\put(0,6){\makebox(12,1){$\tt go \leftarrow p(X)$}}
\end{picture}
\hfill
\begin{picture}(12,7)
\put(0,0){\makebox(12,1){$\tt r(list(U''')) \leftarrow r(U''')$}}
\put(6,1){\line(0,1){1}}
\put(0,2){\makebox(12,1){$\tt r(list(U'')) \leftarrow r(U'')$}}
\put(6,3){\line(0,1){1}}
\put(0,4){\makebox(12,1){$\tt p(list(int)) \leftarrow r(list(int))$}}
\put(6,5){\line(0,1){1}}
\put(0,6){\makebox(12,1){$\tt go \leftarrow p(list(int))$}}
\end{picture}
\caption{A skeleton and the corresponding 
{\em non-proper} type skeleton\label{proper-nonproper-fig} 
for Ex.~\ref{oper-SR-only-ex}}
\end{figure}
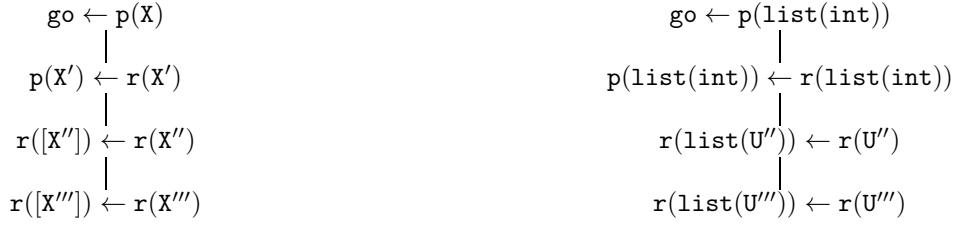

\begin{example}\label{append-ex}
Figure \ref{proper-nonproper-fig} shows a proper skeleton and the
corresponding {\em non-proper} type skeleton for the program in 
Ex.~\ref{oper-SR-only-ex}.

In contrast,
let $\mathcal K$ and $\mathcal F$ be as in 
Ex.~\ref{oper-SR-only-ex}, and 
$\mathcal P = 
\{ \tt app_{list(U),list(U),list(U)},$
$\tt r_{list(int)} \}$. Let $P$ be the program shown in Fig.\
\ref{append-prog-fig}. 
The most general type of each clause is indicated as comment.
Figure~\ref{append-fig} shows a skeleton $S$ and the corresponding
type skeleton $\mathit{TS}$ for $P$.  A solution of 
$\mathit{Eq}(\mathit{TS})$ is obtained
by instantiating all parameters with $\tt int$.
\end{example}

\begin{figure}[t]
\begin{verbatim}
app([],Ys,Ys).                   %app(list(U),list(U),list(U))
app([X|Xs],Ys,[X|Zs]) <-         %app(list(U),list(U),list(U))
  app(Xs,Ys,Zs).                 %app(list(U),list(U),list(U))

r([1]).                          %r(list(int))

go <- 
  app(Xs,[],Zs),                 %app(list(int),list(int),list(int))
  r(Xs).                         %r(list(int)) 
\end{verbatim} 
\caption{A program used to illustrate type skeletons\label{append-prog-fig}}
\end{figure}

\setlength{\unitlength}{0.46cm}
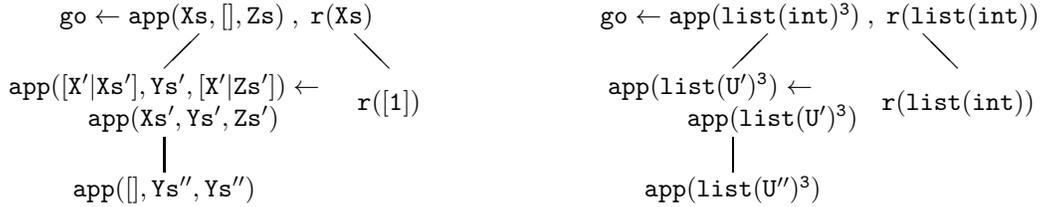
\begin{figure}[t]
\begin{picture}(13,6)
\put(0,0){\makebox(10,1){$\tt app([],Ys'',Ys'')$}}
\put(5,1){\line(0,1){1}}
\put(0,2){\makebox(10,2){$
  \begin{array}{l}
  \tt app([X'|Xs'],Ys',[X'|Zs'])  \leftarrow\\
   \quad \quad \quad \tt app(Xs',Ys',Zs')
  \end{array}$}}
\put(10,2){\makebox(3,2){$\tt r([1])$}}
\put(5,4){\line(1,1){1}}
\put(11.5,4){\line(-1,1){1}}
\put(0,5){\makebox(13,1){$\tt 
  go \leftarrow app(Xs,[],Zs)\; \und \; r(Xs)$}}
\end{picture}
\hfill
\begin{picture}(13,6)
\put(0,0){\makebox(8,1){$\tt app(list(U'')^3)$}}
\put(4,1){\line(0,1){1}}
\put(0,2){\makebox(8,2){$
  \begin{array}{l}
  \tt app(list(U')^3)  \leftarrow\\
   \quad \quad \quad \tt app(list(U')^3)
  \end{array}$}}
\put(8,2){\makebox(5,2){$\tt r(list(int))$}}
\put(4,4){\line(1,1){1}}
\put(10.5,4){\line(-1,1){1}}
\put(0,5){\makebox(13,1){$\tt 
  go \leftarrow app(list(int)^3)\; \und \; r(list(int))$}}
\end{picture}
\caption{A skeleton and the corresponding type skeleton
for Ex.~\ref{append-ex} \label{append-fig}}
\end{figure}

\section{Conditions for Subject Reduction}
\label{conditions-sec}
By Thm.~\ref{type-tree-theorem}, a program has subject reduction
if for each proper skeleton, the corresponding type skeleton is
also proper. A more general sufficient condition consists in ensuring
that {\em any} type skeleton is proper. We call this property 
{\bf type unifiability}. Arguably, type unifiability is in the spirit 
of prescriptive typing, since subject reduction should be independent
of the unifiability of terms, i.e., success or failure of the computation. 
However this view has been challenged in the context of higher-order 
logic programming~\cite{NP92}. 

We conjecture that both subject reduction and type unifiability are
undecidable. Proving this is a topic for future work.

\subsection{The Head Condition}

The head condition is the standard way~\cite{HT92-new} of ensuring type unifiability.

\begin{defi}
\label{HC-def}
A clause 
$C = 
p_{\bar{\tau}}(\bar{t}) \leftarrow 
B$
fulfills the {\bf head condition} if its most general type has the form
$(\bar{\tau},\dots)$.
\end{defi}

Note that by the typing rules in Table~\ref{rules-tab}, clearly the
most general type of $C$ must be $(\bar{\tau},\dots)\Theta$ for some
type substitution $\Theta$. Now the head condition states that
the type of the head arguments must be the declared type of the
predicate, or in other words, $\Theta\restr{\bar{\tau}}=\emptyset$.  
It has been shown previously that typed programs
fulfilling the head condition have operational subject
reduction~\cite[Theorem 1.4.7]{HT92-new}.
By Thm.~\ref{decl-oper-theo}, this means that they have subject
reduction.

\subsection{Generalising the Head Condition}
To reason about the existence of a solution for the equation set of a
type skeleton, we give a sufficient condition for unifiability of a finite set
of term equations. 

\begin{propo}\label{eq-unif}
Let 
$E = \{l_1=r_1,\dots,l_m=r_m\}$ be a set of oriented equations, and
assume an order relation on the equations such that 
$l_1=r_1\rightarrow l_2=r_2$ if $r_1$ and
$l_2$ share a variable. $E$ is unifiable if 
\begin{enumerate}
\item \label{equ-c0} for all $1\leq i < j \leq m$, $r_i$ and $r_j$
have no variable in common,  and
\item \label{equ-c1} the graph of $\rightarrow$ is a partial order, and
\item \label{equ-c2} for all $i\in\onetom$, $l_i$ is an instance of $r_i$.
\end{enumerate}
\end{propo}

In fact, the head condition ensures that $\mathit{Eq}(\mathit{TS})$
meets the above conditions for any type skeleton $\mathit{TS}$. The
equations in $\mathit{Eq}(\mathit{TS})$ have the form 
$p(\bar{\tau}_a) = p(\bar{\tau}_h)$, where $\bar{\tau}_a$ 
is the type of an atom and $\bar{\tau}_h$ is the type of a
head. Taking into account that the ``type clauses'' used for
constructing the equations are renamed apart, all the head types
(r.h.s.) have no parameter in common, the graph of $\rightarrow$ is a
tree isomorphic to $\mathit{TS}$, and, by the head
condition, $\bar{\tau}_a$ is an instance of $\bar{\tau}_h$.
In the next subsection, we show that by decomposing each equation 
$p(\bar{\tau}_a) = p(\bar{\tau}_h)$, one can refine this condition.

\subsection{Semi-generic Programs}\label{semi-generic-subsec}
In the head condition, all arguments of a predicate in clause head
position are ``generic'' (i.e.~their type is the declared type). One
might say that all arguments are ``head-generic''. It is thus possible to
generalise the head condition by partitioning the arguments of each
predicate into those which stay head-generic
and those which one requires to be generic for body atoms. 
The latter ones will be called {\em
body-generic}. 
If we place the head-generic arguments of a
clause head and the body-generic arguments of a clause body on
the right hand sides of the equations associated with a type
skeleton, then Condition~\ref{equ-c2} in Prop.\ \ref{eq-unif} is met.

The other two conditions can be obtained in various ways, more or less
complex to verify (an analysis of the analogous problem of
not being subject to occur check (NSTO) can be found in~\cite{DM93}). 
Taking into account the renaming of
``type clauses'', a relation between two equations amounts to a
shared parameter between a generic argument (r.h.s.) and a non-generic
argument (l.h.s.) of a clause. 
We propose here a condition on the clauses which implies that 
the equations of any skeleton can be ordered.

In the following, an atom written as $p(\bar{s},\bar{t})$
means: $\bar{s}$ and $\bar{t}$ are the vectors of terms 
filling the head-generic and body-generic positions of $p$, 
respectively. The notation $p(\bar{\sigma},\bar{\tau})$, where $\sigma$
and $\tau$ are types, is defined analogously.

\begin{defi}
\label{semi-generic-def}
Let $\_\vdash P\ \mathit{Program}$ and
$\_\vdash C\ \mathit{Clause}$ where
\[
C = 
p_{\bar{\tau}_0,\bar{\sigma}_{m+1}}(\bar{t}_0,\bar{s}_{m+1}) \leftarrow 
p^1_{\bar{\sigma}_1,\bar{\tau}_1}(\bar{s}_1,\bar{t}_1),\dots,
p^m_{\bar{\sigma}_m,\bar{\tau}_m}(\bar{s}_m,\bar{t}_m),
\] 
and 
$\Theta$ the type substitution such that 
$(\bar{\tau}_0,\bar{\sigma}_{m+1},
\bar{\sigma}_1,\bar{\tau}_1,\dots,
\bar{\sigma}_m,\bar{\tau}_m)\Theta$ is the most general type of $C$.
We call $C$ {\bf semi-generic} if
\begin{enumerate}
\item\label{is-linear}
for all $i,j \in \zerotom$, $i \neq j$, 
$\pars(\tau_i\Theta) \cap \pars(\tau_j\Theta) = \emptyset$,
\item\label{is-disjoint}
for all $i \in \onetom$, 
$\pars(\bar{\sigma}_i) \cap 
\bigcup_{i \leq j \leq m} \pars(\bar{\tau}_j)
= \emptyset$, 
\item\label{is-generic}
for all $i \in \zerotom$, $\tau_i \Theta=\tau_i$.
\end{enumerate}

A query $Q$ is {\bf semi-generic} if the 
clause ${\tt go} \leftarrow Q$ is semi-generic.
A program is {\bf semi-generic} if each of its clauses is
semi-generic.
\end{defi}

Note that semi-genericity has a strong resemblance with 
{\em nicely-modedness}, where head-generic corresponds to input, and
body-generic corresponds to output. Nicely-modedness has been used,
among other things, to show that programs are free from 
unification~\cite{AE93}. Semi-genericity serves a very similar purpose
here.  Note also that a typed program which fulfills the head
condition is semi-generic, where all argument positions are
head-generic.  

The following theorem states subject reduction for semi-generic
programs.

\begin{rmtheorem}\label{condgen-thm}
Every semi-generic program $P$ has
subject reduction wrt.\ the set of semi-generic queries.
\end{rmtheorem}
\begin{proof}
Let $Q$ be a semi-generic query and $TS$ a type skeleton corresponding 
to a skeleton  for
$P\cup \{{\tt go} \leftarrow Q\}$ with head $\tt go$. 
Each equation in $Eq(TS)$ originates from a pair of nodes 
$(n,n_i)$ where $n$ is labelled
$C = p(\bar{\tau}_0,\bar{\sigma}_{m+1}) \leftarrow 
p_1(\bar{\sigma}_1,\bar{\tau}_1),\dots,
p_m(\bar{\sigma}_m,\bar{\tau}_m)$ and $n_i$ is labelled
$C_i = p_i(\bar{\tau}'_i,\bar{\sigma}'_i)\leftarrow\dots$, and the equation
is 
$p_i(\bar{\tau}'_i,\bar{\sigma}'_i) =
 p_i(\bar{\sigma}_i,\bar{\tau}_i)$.
Let
$Eq'$ be obtained from $Eq(TS)$ by replacing each such equation with
the two equations
$\bar{\sigma}_i = \bar{\tau}'_i$,
$\bar{\sigma}'_i = \bar{\tau}_i$. 
Clearly $Eq'$ and $Eq(TS)$ are equivalent. Because of the renaming of
parameters for each node and since $TS$ is a tree, it is possible to
define an order $\dashrightarrow$ on the equations in $Eq'$ such that 
for {\em each} label $C$ defined as above, 
$\bar{\sigma}_1 = \bar{\tau}'_1 \dashrightarrow 
E_1 \dashrightarrow 
\bar{\sigma}'_1 = \bar{\tau}_1 \dashrightarrow 
\dots \dashrightarrow
\bar{\sigma}_m = \bar{\tau}'_m \dashrightarrow 
E_m \dashrightarrow 
\bar{\sigma}'_m = \bar{\tau}_m$, 
where for each $i\in\onetom$, 
$E_i$ denotes a sequence containing all equations
$e$ with $\pars (e) \cap \pars(C_i) \neq \emptyset$.

We show that $Eq'$ fulfills the conditions of Prop.\ \ref{eq-unif}. 
By Def.~\ref{semi-generic-def} (\ref{is-linear}), $Eq'$ fulfills
condition~\ref{equ-c0}.
By  Def.~\ref{semi-generic-def} (\ref{is-disjoint}), it follows that 
$\rightarrow$ is a subrelation of $\dashrightarrow$, and hence 
$Eq'$ fulfills condition~\ref{equ-c1}.
By Def.~\ref{semi-generic-def} (\ref{is-generic}), $Eq'$ fulfills
condition~\ref{equ-c2}. 

Thus $Eq(TS)$ has a solution, so $TS$ is proper, and so by Thm.\
\ref{type-tree-theorem}, $P$ has subject reduction wrt.\ the set of
semi-generic queries. 
\end{proof}

The following example shows that our condition
extends the class of programs that have subject reduction. 

\begin{example}\label{semi-generic-ex}
Suppose $\mathcal{K}$ and $\mathcal{F}$ define lists as usual (see
Ex.\ \ref{oper-SR-only-ex}). Let 
$\mathcal{P} = 
\{\tt p_{U,V}, q_{U,V}\}$ and assume that for
$\tt p, q$, the first argument is head-generic and the 
second argument is body-generic.
Consider the following program.

\begin{verbatim}
p(X,[Y]) <-               %p(U,list(V)) <-
  q([X],Z), q([Z],Y).     %  q(list(U),W), q(list(W),V).
q(X,[X]).                 %q(U,list(U)).
\end{verbatim}

This program is semi-generic. E.g.\ in the first type clause the terms
in generic positions are $\tt U,W, V$; all generic arguments have the
declared type (condition~\ref{is-generic}); they do not share a
parameter (condition~\ref{is-linear}); no generic argument in the body
shares a parameter with a non-generic position to the left of it
(condition~\ref{is-disjoint}).  A type skeleton is shown in
Fig.~\ref{semi-generic-fig}.

\setlength{\unitlength}{0.52cm}
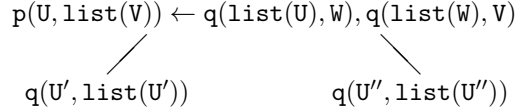
\begin{figure}[t]
\begin{center}
\begin{picture}(13,3)
\put(0,0){\makebox(5,1){$\tt q(U',list(U'))$}}
\put(8,0){\makebox(5,1){$\tt q(U'',list(U''))$}}
\put(2.5,1){\line(1,1){1}}
\put(10.5,1){\line(-1,1){1}}
\put(0,2){\makebox(13,1){$
  \tt p(U,list(V)) \leftarrow q(list(U),W), q(list(W),V)$}}
\end{picture}
\end{center}
\caption{A type skeleton 
for a semi-generic program\label{semi-generic-fig}}
\end{figure}

As another example, suppose now that 
$\mathcal{K}$ and $\mathcal{F}$ define list and integers, 
and consider the predicate ${\tt r}/2$ specified as
$\tt r(1,[]), r(2,[[]]), r(3,[[[]]])\dots$. Its obvious definition
would be

{
\begin{verbatim}
r(1,[]).
r(J,[X]) <-  r(J-1,X).
\end{verbatim}
} 
One can see that this program must violate the head condition no
matter what the declared type of $\tt r$ is. However, assuming
declared type $\tt (int,list(U))$ and letting the second argument be
body-generic, the program is semi-generic.
\end{example}

One can argue that in the second example, there is an intermingling of 
the typing and the computation, which contradicts the spirit of
prescriptive typing. However, as we discuss in the next section,
the situation is not always so clearcut.


\section{What is the Use of the Head Condition?}
\label{is-restrictive-sec}
The above results shed new light on the head condition.  They allow us
to view it as just one particularly simple condition guaranteeing
type unifiability and consequently subject reduction and
``well-typing'' of the result, and hence a certain correctness of the
program. This raises the question whether by generalising the
condition, we have significantly enlarged the class of ``well-typed''
programs.

However, the head condition is also sometimes viewed as a condition
inherent in the type system, or more specifically, an essential
characteristic of {\em generic} polymorphism, as opposed to 
{\em ad-hoc} polymorphism. Generic polymorphism means that
predicates are defined on an infinite number of types and that the
definition is independent of a particular instance of the
parameters. Ad-hoc polymorphism, often called 
{\em overloading}~\cite{M78}, means, e.g., to use the same
symbol $+$ for integer addition, matrix addition and list
concatenation.  Ad-hoc polymorphism is in fact forbidden by the head
condition.  

One way of reconciling ad-hoc polymorphism with the head condition is
to enrich the type system so that types can be passed as parameters,
and the definition of a predicate depends on these
parameters~\cite{LR96}.  
Under such conditions, the head condition is regarded as natural.

So as a second, more general question, 
we discuss the legitimacy of the head condition
briefly, since the answer justifies the interest 
in our first question.

In favour of the head condition, one could argue (1) that a program typed
in this way does not compute types, but only propagates them; (2) that it
allows for separate compilation since an imported predicate can be
compiled without consulting its definition; and (3) that it disallows
certain ``unclean'' programs~\cite{craft}. 

In reality, these points are not, strictly speaking, fundamental
arguments in favour of the head condition. Our generalisation does not
necessarily imply a confusion between computation and typing (even if
the result type does not depend on the result of a computation, it may
be an instance of the declared type). Moreover, if the type
declarations of the predicates are accompanied by declarations of the
head- and body-generic arguments, separate compilation remains
possible. Finally, Hanus~\cite{Han92} does not consider the head
condition to be particularly natural, arguing that it is an
important feature of logic programming that it allows for 
{\em lemma generation}. 

We thus believe that the first question is, after all, relevant. So
far, we have not been able to identify a ``useful'', non-contrived,
example which clearly shows the interest in the class of
semi-generically typed programs. The following example demonstrates
the need for a generalisation, but also the insufficiency of the class
defined in Def.~\ref{semi-generic-def}.

\begin{example}\label{fg-ex}
Let
$\mathcal{K} =$
$\{\mathtt{t}/1, \mathtt{int}/0\}$ and
\[
\mathcal{F} = 
\{\mathtt{-1_{\rightarrow int}}, 
\mathtt{0_{\rightarrow int}},\dots,
\mathtt{c}_\mathtt{\rightarrow t(U)},
\mathtt{g}_\mathtt{U \rightarrow t(U)},
\mathtt{f}_\mathtt{t(t(U)) \rightarrow t(U)}\}.
\]
For all 
$i \geq 0$, we have
$\_\vdash \mathtt{g}^i({\tt c}) : {\tt t}^{i+1}({\tt U})$ and 
$\_\vdash \mathtt{f}^i(\mathtt{g}^i({\tt c})) : {\tt t}({\tt U})$. 
This means that the set 
$
\{\sigma \mid \exists s,t.\ \mbox{$s$ is subterm of $t$}, 
                          \ \mbox{$\_\vdash s:\sigma$},
                          \ \mbox{$\_\vdash t:{\tt t(U)}$} \}
$ 
is infinite, or in words, there are infinitely many types that a
subterm of a term of type $\tt t(U)$ can have. 
This property of the type $\tt t(U)$ 
is very unusual. In~\cite{SHK00}, a condition is considered (the 
{\em Reflexive Condition}) which rules out this situation.

Now consider the predicate ${\tt fgs}/2$ specified as 
${\tt fgs}(i,\mathtt{f}^i(\mathtt{g}^i({\tt c})))$ ($i \in \nat$).
Figure~\ref{fg-fig} 
presents three potential definitions of this
predicate. The declared types of the predicates are given by
$\mathcal{P} = $
$\{\tt fgs1_{int, t(U)}, $
$\tt gs1_{int, t(U)},$ 
$\tt fgs2_{int, t(U)},$ 
$\tt fgs3_{int, t(U)},$
$\tt fs1_{int, t(U), int},$ 
$\tt fs2_{int, t(U), int},$\linebreak
$\tt gs2_{int, t(U), t(V)},$ 
$\tt fgs3\_aux_{int, t(U), t(U)}\}$.
The first solution is the most straightforward one, but its
last clause does not fulfill the head condition.  For the second
solution, the fact clause \verb@gs2(0,x,x).@ does not
fulfill the head condition. The third program fulfills the head
condition but is the least obvious solution.

\begin{figure}[t]
\begin{minipage}{2.95cm}
\small
\begin{verbatim}
fgs1(I,Y) <- 
  fs1(I,Y,I).

fs1(I,f(X),J) <-
  fs1(I-1,X,J).
fs1(0,X,J) <-
  gs1(J,X).

gs1(J,g(X)) <-
  gs1(J-1,X).
gs1(0,c).
\end{verbatim}
\end{minipage}
\hfill
\vline
\hfill
\begin{minipage}{3.35cm}
\small
\begin{verbatim}
fgs2(I,Y) <- 
  fs2(I,Y,I).
 
fs2(I,f(X),J) <-
  fs2(I-1,X,J).
fs2(0,X,J) <-
  gs2(J,X,c).

gs2(J,X,Y) <-
  gs2(J-1,X,g(Y)).
gs2(0,X,X).
\end{verbatim}
\end{minipage}
\hfill
\vline
\hfill
\begin{minipage}{4.3cm}
\small
\begin{verbatim}
fgs3(I,X) <-
  fgs3_aux(I,c,X).

fgs3_aux(I,X,f(Y)) <-
  fgs3_aux(I-1,g(X),Y).
fgs3_aux(0,X,X).





\end{verbatim}
\end{minipage}
\caption{Three potential solutions for Ex.~\ref{fg-ex}\label{fg-fig}}
\end{figure}
\end{example}

For the above example, the head condition is a real restriction. 
It prevents a solution using
the most obvious algorithm, which is certainly a drawback of any type
system. We suspected initially that it would be impossible
to write a program fulfilling the specification of $\tt fgs$ without
violating the head condition. 

Now it would of course be interesting to see if the first two
programs, which violate the head condition, are
semi-generic. Unfortunately, they are not. We explain this for the
first program. The second position of 
\verb@gs1@ must be body-generic because of the second clause for
\verb@gs1@.  This implies that the second position of 
\verb@fs1@ must also be body-generic because of the second clause
for \verb@fs1@ 
(otherwise there would be two generic positions with a common
parameter). That however is unacceptable for the first clause of 
\verb@fs1@ ($\tt X$ has type $\tt t(t(U))$, instance of 
$\tt t(U)$). 

It can however be observed that both programs have subject
reduction wrt.\ the queries 
${\tt fgs}j(i,{\tt Y})$ for $i \in \nat$ and $j = 1,2$. 
In fact for these queries all type
skeletons are proper, but it can be seen that the
equations associated with the type skeletons cannot be ordered.  
This shows that the condition of semi-genericity is still too
restrictive.

There is a perfect analogy between \verb@gs1@ and $\tt r$ in Ex.\
\ref{semi-generic-ex}.

To conclude this section, note that our solution to the problem in
Ex.\ \ref{fg-ex} uses {\em polymorphic recursion}, a concept
previously discussed for functional programming~\cite{KFU93-short}:
In the recursive clause for {\tt fgs3\_aux}, the arguments of the 
recursive call have type 
$\tt (int,t(t(U)),t(t(U)))$, 
while the arguments of the clause head have type
$\tt (int,t(U),t(U))$. 
If we wrote a function corresponding to \verb@fgs3_aux@ in 
Miranda~\cite{Tho95} or ML, the type checker
could not infer its type, since it assumes that recursion is
monomorphic, i.e., the type of a recursive call is identical to the
type of the ``head''. In Miranda, this problem can be overcome by 
providing a type declaration, while in ML, the function will 
definitely be rejected. This limitation of the ML type system, or 
alternatively, the ML type checker, has been studied by Kahrs~\cite{K96}. 

There is a certain duality between the head condition and monomorphic
recursion. When trying to find a solution to our problem, we found
that we either had to violate the head condition or use polymorphic
recursion.
For example, in the recursive clause for {\tt gs1}, 
the arguments of the recursive call have type 
$\tt (int,t(U))$, 
while the arguments of the clause head have type
$\tt (int,t(t(U)))$, which is in a way the reverse of the situation
for  \verb@fgs3_aux@. Note that this implies a violation of the
head condition for {\em any} declared type of {\tt gs1}. It would be
interesting to investigate this duality further.

\section{Conclusion}\label{conclusion-sec}

In this paper we redefined the notion of {\em subject
reduction} by using derivation trees, leading to a
proof-theoretic view of typing in logic programming.
We showed that this new notion is equivalent to the operational
one (Thm.\ \ref{decl-oper-theo}).

We introduced {\em type skeletons}, obtained from skeletons by
replacing terms with their types. We showed that a program has subject
reduction if for each proper skeleton, the type skeleton
is also proper. Apart from clarifying the motivations of the 
head condition, it has several potential applications:

\begin{itemize}
\item 
It facilitates studying the
semantics of typed programs by simplifying its formulation in
comparison to other works (e.g.\ \cite{LR91}). Lifting the notions of
derivation tree and skeleton on the level of types can  
help formulate proof-theoretic and operational semantics,
just as this has
been done for untyped logic programming with the classical trees
\cite{BGLM91,DM93,FLMP89}.
\item 
The approach may enhance program analysis based on abstract
interpretation. 
Proper type skeletons could also be modelled
by fixpoint operators \cite{CominiLMV96,CC77,GDL95}. 
Abstract interpretation for prescriptively typed programs has been
studied by~\cite{RBM99,SHK00}, and it has been pointed out that the
head condition is essential for ensuring that the abstract semantics
of a program is finite, which is crucial for the termination of an
analysis. It would be interesting to investigate the impact of more
general conditions. 
\item
This ``proof-theoretic'' approach to typing could also be applied for
synthesis of typed programs. In~\cite{TBD97}, the authors propose 
the automatic generation of lemmas, using synthesis techniques based
on resolution. It is interesting to observe that the generated lemmas 
meet the head condition, which our approach seems to be able to
justify and even generalise.
\item 
The approach may help in combining {\em prescriptive} and 
{\em descriptive} approaches to typing. The latter are usually based
on partial correctness properties.  Descriptive type systems satisfy
certain criteria of type-correctness \cite{DM98}, but subject
reduction is difficult to consider in such systems. Our approach is a
step towards potential combinations of different approaches.
\end{itemize}

We have presented a condition for type unifiability which is a
refinement of the head condition (Thm.~\ref{condgen-thm}). Several
observations arise from this:
\begin{itemize}
\item
Definition~\ref{semi-generic-def} is decidable. If the partitioning of
the arguments is given, it can be verified in polynomial
time. Otherwise, finding a partitioning is exponential in the number
of argument positions.

\item 
The refinement has a cost: subject reduction does not hold for
arbitrary (typed) queries. The head condition, by its name, only
restricts the clause heads, whereas our generalisation also restricts
the queries, and hence the ways in which a program can be used.

\item As we have seen, the proposed refinement may not be
sufficient. Several approaches can be used to introduce further
refinements based on abstract interpretation or on properties of sets
of equations. Since any sufficient condition for type unifiability
contains at least an NSTO condition, one could also benefit from
the refinements proposed for the NSTO check \cite{DM93}.  
Such further refined conditions should, in particular, be fulfilled by
all solutions of Ex.~\ref{fg-ex}.
\end{itemize}

We have also studied {\em operational} subject reduction for type
systems with subtyping~\cite{SFD00}.  As future work, we want to
integrate that work with the {\em proof-theoretic} view of subject
reduction of this paper. Also, we want to prove the undecidability of
subject reduction and type unifiability, and design more refined tests 
for type unifiability.

\subsection*{Acknowledgements}
We thank Fran\c{c}ois Fages for interesting discussions. Jan-Georg
Smaus was supported by an ERCIM fellowship.

{\small
\bibliography{thesis,modes_types}

\begin{thebibliography}{10}

\bibitem{AE93}
K.~R. Apt and S.~Etalle.
\newblock On the unification free {P}rolog programs.
\newblock In A.~Borzyszkowski and S.~Sokolowski, editors, {\em Proceedings of
  the Conference on Mathematical Foundations of Computer Science}, volume 711
  of {\em LNCS}, pages 1--19. Springer-Verlag, 1993.

\bibitem{B95}
C.~Beierle.
\newblock Type inferencing for polymorphic order-sorted logic programs.
\newblock In L.~Sterling, editor, {\em Proceedings of the Twelfth International
  Conference on Logic Programming}, pages 765--779. MIT Press, 1995.

\bibitem{BGLM91}
A.~Bossi, M.~Gabbrielli, G.~Levi, and M.~Martelli.
\newblock The $s$-semantics approach: theory and applications.
\newblock {\em Journal of Logic Programming}, 19/20:149--197, 1991.

\bibitem{CominiLMV96}
M.~Comini, G.~Levi, M.~C. Meo, and G.~Vitiello.
\newblock Proving properties of logic programs by abstract diagnosis.
\newblock In M.~Dams, editor, {\em Analysis and Verification of Multiple-Agent
  Languages, 5th LOMAPS Workshop}, volume 1192 of {\em LNCS}, pages 22--50.
  Springer-Verlag, 1996.

\bibitem{CC77}
P.~Cousot and R.~Cousot.
\newblock Abstract interpretation: {A} unified lattice model for static
  analysis of programs by construction or approximation of fixpoints.
\newblock In {\em Proceedings of the 4th Symposium on Principles of Programming
  Languages}, pages 238--252. ACM Press, 1977.

\bibitem{DM93}
P.~Deransart and J.~Ma{\l}uszy\'nski.
\newblock {\em A Grammatical View of Logic Programming}.
\newblock MIT Press, 1993.

\bibitem{DM98}
P.~Deransart and J.~Ma{\l}uszy\'nski.
\newblock Towards soft typing for {CLP}.
\newblock In F.~Fages, editor, {\em JICSLP'98 Post-Conference Workshop on Types
  for Constraint Logic Programming}. {\'E}cole Normale Sup{\'e}rieure, 1998.
\newblock Available at {\tt http://discipl.inria.fr/TCLP98/}.

\bibitem{DS01}
P.~Deransart and J.-G. Smaus.
\newblock Well-typed logic programs are not wrong.
\newblock In H.~Kuchen and K.~Ueda, editors, {\em Proceedings of the 5th
  International Symposium on Functional and Logic Programming}, LNCS.
  Springer-Verlag, 2001.

\bibitem{FLMP89}
M.~Falaschi, G.~Levi, M.~Martelli, and C.~Palamidessi.
\newblock Declarative modeling of the operational behavior of logic languages.
\newblock {\em Theoretical Computer Science}, 69(3):289--318, 1989.

\bibitem{GDL95}
R.~Giacobazzi, S.~K. Debray, and G.~Levi.
\newblock Generalized semantics and abstract interpretation for constraint
  logic programs.
\newblock {\em Journal of Logic Programming}, 25(3):191--247, 1995.

\bibitem{Han92}
M.~Hanus.
\newblock {\em Logic Programming with Type Specifications}, chapter~3, pages
  91--140.
\newblock In~\cite{P92}.

\bibitem{goedel}
P.~M. Hill and J.~W. Lloyd.
\newblock {\em {The G{\"o}del Programming Language}}.
\newblock MIT Press, 1994.

\bibitem{HT92-new}
P.~M. Hill and R.~W. Topor.
\newblock {\em A Semantics for Typed Logic Programs}, chapter~1, pages 1--61.
\newblock In~\cite{P92}.

\bibitem{K96}
S.~Kahrs.
\newblock Limits of {M}{L}-definability.
\newblock In H.~Kuchen and S.~D. Swierstra, editors, {\em Proceedings of the
  8th Symposium on Programming Language Implementations and Logic Programming},
  volume 1140 of {\em LNCS}, pages 17--31. Springer-Verlag, 1996.

\bibitem{KFU93-short}
A.~J. Kfoury, J.~Tiuryn, and P.~Urzyczyn.
\newblock Type reconstruction in the presence of polymorphic recursion.
\newblock {\em ACM Transactions on Programming Languages and Systems},
  15(2):290--311, 1993.

\bibitem{LR91}
T.K. Lakshman and U.S. Reddy.
\newblock Typed {P}rolog: A semantic reconstruction of the
  {M}ycroft-{O}'{K}eefe type system.
\newblock In V.~Saraswat and K.~Ueda, editors, {\em Proceedings of the 1991
  International Symposium on Logic Programming}, pages 202--217. MIT Press,
  1991.

\bibitem{L87}
J.~W. Lloyd.
\newblock {\em Foundations of Logic Programming}.
\newblock Springer-Verlag, 1987.

\bibitem{LR96}
P.~Louvet and O.~Ridoux.
\newblock Parametric polymorphism for {T}yped {P}rolog and $\lambda${P}rolog.
\newblock In H.~Kuchen and S.~D. Swierstra, editors, {\em Proceedings of the
  8th Symposium on Programming Language Implementations and Logic Programming},
  volume 1140 of {\em LNCS}, pages 47--61. Springer-Verlag, 1996.

\bibitem{MM82}
A.~Martelli and U.~Montanari.
\newblock An efficient unification algorithm.
\newblock {\em ACM Transactions on Programming Languages and Systems},
  4:258--282, 1982.

\bibitem{M78}
R.~Milner.
\newblock A theory of type polymorphism in programming.
\newblock {\em Journal of Computer and System Sciences}, 17(3):348--375, 1978.

\bibitem{MO84}
A.~Mycroft and R.~O'Keefe.
\newblock A polymorphic type system for {P}rolog.
\newblock {\em Artificial Intelligence}, 23:295--307, 1984.

\bibitem{NP92}
G.~Nadathur and F.~Pfenning.
\newblock {\em Types in Higher-Order Logic Programming}, chapter~9, pages
  245--283.
\newblock In \cite{P92}.

\bibitem{craft}
R.~A. O'Keefe.
\newblock {\em {The Craft of {P}rolog}}.
\newblock MIT Press, 1990.

\bibitem{P92}
F.~Pfenning, editor.
\newblock {\em Types in Logic Programming}.
\newblock MIT Press, 1992.

\bibitem{RBM99}
O.~Ridoux, P.~Boizumault, and F.~Mal\'esieux.
\newblock Typed static analysis: Application to groundness analysis of {P}rolog
  and $\lambda${P}rolog.
\newblock In A.~Middeldorp and T.~Sato, editors, {\em Proceedings of the 4th
  Fuji International Symposium on Functional and Logic Programming}, volume
  1722 of {\em LNCS}, pages 267--283. Springer-Verlag, 1999.

\bibitem{SFD00}
J.-G. Smaus, F.~Fages, and P.~Deransart.
\newblock Using modes to ensure subject reduction for typed logic programs with
  subtyping.
\newblock In S.~Kapoor and S.~Prasad, editors, {\em Proceedings of the 20th
  Conference on the Foundations of Software Technology and Theoretical Computer
  Science}, volume 1974 of {\em LNCS}. Springer-Verlag, 2000.

\bibitem{SHK00}
J.-G. Smaus, P.~M. Hill, and A.~M. King.
\newblock Mode analysis domains for typed logic programs.
\newblock In A.~Bossi, editor, {\em Proceedings of the 9th International
  Workshop on Logic-based Program Synthesis and Transformation}, volume 1817 of
  {\em LNCS}, pages 83--102, 2000.

\bibitem{mercury}
Z.~Somogyi, F.~Henderson, and T.~Conway.
\newblock The execution algorithm of {Mercury}, an efficient purely declarative
  logic programming language.
\newblock {\em Journal of Logic Programming}, 29(1--3):17--64, 1996.

\bibitem{TBD97}
P.~Tarau, K.~De~Bosschere, and B.~Demoen.
\newblock On {D}elphi lemmas and other memoing techniques for deterministic
  logic programs.
\newblock {\em Journal of Logic Programming}, 30(2):145--163, 1997.

\bibitem{Tho91}
Simon Thompson.
\newblock {\em Type Theory and Functional Programming}.
\newblock Addison-Wesley, 1991.

\bibitem{Tho95}
Simon Thompson.
\newblock {\em Miranda: The Craft of Functional Programming}.
\newblock Addison-Wesley, 1995.

\end{thebibliography}
}

\tableofcontents
\end{document}